\documentclass{elsart}
\usepackage{graphics}
\usepackage{epsfig}

\usepackage{amsmath}

\newcommand{\aff}[2]{Dipartimento di Fisica dell'Universit\`a #1 e Sezione INFN, #2, Italy.}
\newcommand{\affd}[1]{Dipartimento di Fisica dell'Universit\`a e Sezione INFN, #1, Italy.}

\def\DAF{DA\char8NE}  \def\epm{\ifm{e^+e^-}}
\def\ifm#1{\relax\ifmmode#1\else$#1$\fi}

\def\ff{$\phi$--factory}

  \def\x{\ifm{\times}}
\def\pt#1,#2,{\ifm{#1\x10^{#2}}}
\newcommand{\kl}{\mbox{$K_L$}}
\newcommand{\ks}{\mbox{$K_S$}}

\newcommand{\Pphi}{\ensuremath{\phi}}

\newcommand{\eV}{{e\kern-.07em V}}
\newcommand{\kloe}{K{\kern-.07em LOE} }
\newcommand{\dafne}{D{\kern-.07em A\ensuremath{\Phi}NE }}
\newcommand{\keV}{{\rm \,k\eV}}
\newcommand{\MeV}{{\rm \,M\eV}}
\newcommand{\GeV}{{\rm G\eV}}
\newcommand{\ps}{{\rm \,ps}}
\newcommand{\ns}{{\rm \,ns}}
\newcommand{\mm}{{\rm \,mm}}
\newcommand{\cm}{{\rm \,cm}}
\newcommand{\m}{{\rm \,m}}
\newcommand{\mrad}{{\rm \,mrad}}

\newcommand{\um}{\ensuremath{\mathrm{\mu m}}}
\newcommand{\T}{{\rm \,T}}
\newcommand{\Lnb}{\ensuremath{\rm \, nb^{-1}}}

\newcommand{\Lfb}{\ensuremath{\rm \, fb^{-1}}}

\newcommand{\DKSpippim}{\ensuremath{K_S\rightarrow\pi^+\pi^-}}

\newcommand{\DKSee}{\ensuremath{K_S\rightarrow e^+e^-}}
\newcommand{\DKLee}{\ensuremath{K_L\rightarrow e^+e^-}}

\newcommand{\DKLmm}{\ensuremath{K_L\rightarrow \mu^+\mu^-}}
\newcommand{\DKSgg}{\ensuremath{K_S\rightarrow \gamma \gamma}}

\newcommand{\Dphipippimpio}{\ensuremath{\phi\rightarrow\pi^+\pi^-\pi^0}}

\newcommand{\BR}[1]{\ensuremath{\mathrm{BR}(#1)}}

\newcommand{\kcr}{\ensuremath{K_\mathrm{crash}}}

\newcommand{\koko}{\ensuremath{K^0\bar{K}^0}}

\begin{document}

\begin{frontmatter}
\title{Search for the $\DKSee$ decay with the $\kloe$ detector at $\DAF$}
\collab{The KLOE Collaboration}
\author[Na]{F.~Ambrosino},
\author[Frascati]{A.~Antonelli},
\author[Frascati]{M.~Antonelli},
\author[Frascati]{F.~Archilli\corauthref{cor1}},
\author[Roma3]{C.~Bacci},
\author[Karlsruhe]{P.~Beltrame},
\author[Frascati]{G.~Bencivenni},
\author[Frascati]{S.~Bertolucci},
\author[Roma1]{C.~Bini},
\author[Frascati]{C.~Bloise},
\author[Roma3]{S.~Bocchetta},
\author[Roma1]{V.~Bocci},
\author[Frascati]{F.~Bossi},
\author[Roma3]{P.~Branchini},
\author[Roma1]{R.~Caloi},
\author[Frascati]{P.~Campana},
\author[Frascati]{G.~Capon},
\author[Na]{T.~Capussela},
\author[Roma3]{F.~Ceradini},
\author[Frascati]{S.~Chi},
\author[Na]{G.~Chiefari},
\author[Frascati]{P.~Ciambrone},
\author[Frascati]{E.~De~Lucia},
\author[Roma1]{A.~De~Santis},
\author[Frascati]{P.~De~Simone},
\author[Roma1]{G.~De~Zorzi},
\author[Karlsruhe]{A.~Denig},
\author[Roma1]{A.~Di~Domenico},
\author[Na]{C.~Di~Donato},
\author[Pisa]{S.~Di~Falco},
\author[Roma3]{B.~Di~Micco},
\author[Na]{A.~Doria},
\author[Frascati]{M.~Dreucci},
\author[Frascati]{G.~Felici},
\author[Frascati]{A.~Ferrari},
\author[Frascati]{M.~L.~Ferrer},
\author[Frascati]{G.~Finocchiaro},
\author[Roma1]{S.~Fiore},
\author[Frascati]{C.~Forti},
\author[Roma1]{P.~Franzini},
\author[Frascati]{C.~Gatti},
\author[Roma1]{P.~Gauzzi},
\author[Frascati]{S.~Giovannella},
\author[Lecce]{E.~Gorini},
\author[Roma3]{E.~Graziani},
\author[Pisa]{M.~Incagli},
\author[Karlsruhe]{W.~Kluge},
\author[Moscow]{V.~Kulikov},
\author[Roma1]{F.~Lacava},
\author[Frascati]{G.~Lanfranchi},
\author[Frascati,StonyBrook]{J.~Lee-Franzini},
\author[Karlsruhe]{D.~Leone},
\author[Frascati]{M.~Martini},
\author[Na]{P.~Massarotti},
\author[Frascati]{W.~Mei},
\author[Na]{S.~Meola},
\author[Frascati]{S.~Miscetti},
\author[Frascati]{M.~Moulson},
\author[Frascati]{S.~M\"uller},
\author[Frascati]{F.~Murtas},
\author[Na]{M.~Napolitano},
\author[Roma3]{F.~Nguyen},
\author[Frascati]{M.~Palutan\corauthref{cor2}},
\author[Roma1]{E.~Pasqualucci},
\author[Roma3]{A.~Passeri},
\author[Frascati,Energ]{V.~Patera},
\author[Na]{F.~Perfetto},
\author[Lecce]{M.~Primavera},
\author[Frascati]{P.~Santangelo},
\author[Na]{G.~Saracino},
\author[Frascati]{B.~Sciascia},
\author[Frascati,Energ]{A.~Sciubba},
\author[Pisa]{F.~Scuri},
\author[Frascati]{I.~Sfiligoi},
\author[Frascati]{T.~Spadaro\corauthref{cor3}},
\author[Roma1]{M.~Testa},
\author[Roma3]{L.~Tortora},
\author[Roma1]{P.~Valente},
\author[Karlsruhe]{B.~Valeriani},
\author[Frascati]{G.~Venanzoni},
\author[Frascati]{R.Versaci},
\author[Frascati,Beijing]{G.~Xu}

\address[Frascati]{Laboratori Nazionali di Frascati dell'INFN, 
Frascati, Italy.}
\address[Karlsruhe]{Institut f\"ur Experimentelle Kernphysik, 
Universit\"at Karlsruhe, Germany.}
\address[Lecce]{\affd{Lecce}}
\address[Na]{Dipartimento di Scienze Fisiche dell'Universit\`a 
``Federico II'' e Sezione INFN,
Napoli, Italy}
\address[Pisa]{\affd{Pisa}}
\address[Energ]{Dipartimento di Energetica dell'Universit\`a 
``La Sapienza'', Roma, Italy.}
\address[Roma1]{\aff{``La Sapienza''}{Roma}}
\address[Roma3]{\aff{``Roma Tre''}{Roma}}
\address[StonyBrook]{Physics Department, State University of New 
York at Stony Brook, USA.}
\address[Beijing]{Permanent address: Institute of High Energy 
Physics of Academica Sinica,  Beijing, China.}
\address[Moscow]{Permanent address: Institute for Theoretical 
and Experimental Physics, Moscow, Russia.}
\corauth[cor1]{Corresponding author: flavio.archilli@lnf.infn.it}
\corauth[cor2]{Corresponding author: matteo.palutan@lnf.infn.it}
\corauth[cor3]{Corresponding author: tommaso.spadaro@lnf.infn.it}
\begin{abstract}
We present results of a direct search for the decay \DKSee\ with the \kloe\ detector, obtained with
a sample of $\epm \to \Pphi \to \ks\kl$ events produced at \DAF, the Frascati \ff, for an integrated luminosity
of 1.3~$\Lfb$. The Standard Model prediction for this decay is $\BR\DKSee = 1.6\times 10^{-15}$.
The search has been performed tagging the \ks\ decays by simultaneous detection of a 
\kl\ interaction in the calorimeter. Background rejection has been optimized by using both kinematic cuts and
particle identification. At the end of the analysis chain we find $\BR\DKSee < 2.1\times10^{-8}$ at 90\%~CL, 
which improves by a factor of $\sim 7$ on the previous best result, obtained by CPLEAR experiment. 
\end{abstract}
\end{frontmatter}

\section{Introduction}
\begin{figure}[h!]
  \begin{center}    
    \includegraphics[totalheight=3.cm]{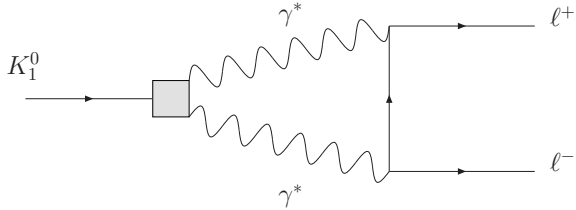}
    \caption{Long distance contribution to $\ks \rightarrow \ell^+ \ell^-$ process, mediated by two-photon exchange.}
    \label{fig:sec1.fig2}
  \end{center}
\end{figure}

The decay \DKSee, like the decay  \DKLee\ or \DKLmm, is a flavour-changing neutral-current process, 
suppressed in the Standard Model and dominated by the two-photon intermediate state~\cite{kseetheory}.
For both \ks\ and\kl, the \epm\ channel is much more suppressed than the $\mu^+\mu^-$ one (by a factor of $\sim 250$)
because of the $e-\mu$ mass difference. 
The diagram corresponding to the process $\ks \rightarrow \gamma^* \gamma^* \rightarrow \ell^+ \ell^-$ is shown in 
Fig.~\ref{fig:sec1.fig2}.
Using Chiral Perturbation Theory  ($\chi$pT) to order $\mathcal{O}(p^4)$, G.~Ecker and A.~Pich evaluated the ratio~\cite{kseetheory}
$\Gamma(K_S \rightarrow e^+ e^-)/\Gamma(K_S \rightarrow \gamma \gamma) = 8 \times 10^{-9}$, with $10\%$ uncertainty.
Using the present average~\cite{pdg06} for BR(\DKSgg) 
we obtain the Standard Model prediction BR$(\DKSee) \simeq 10 ^{-15}$.
A value significantly higher than expected would point to new physics.
The best experimental limit for $BR(\DKSee)$ has been measured
by CPLEAR~\cite{cplearksee}, and it is equal to $1.4\times 10^{-7}$, at $90\%$~CL.
Here we present a new measurement of this channel, which improves on the previous result
by a factor of $\sim 7$.
This paper is organized as follows: in the next section, a brief description of the \kloe\ 
experimental setup is given; in section~\ref{sec:ana} the selection criteria for the decays of
interest are summarized. Results are presented in section~\ref{sec:re}.

\section{Experimental setup}
\label{sec:expsetup}

The data were collected with \kloe\ detector at \DAF, the Frascati \ff. 
\DAF\ is an \epm\ collider that operates at a 
center-of-mass energy of $\sim 1020\MeV$, the mass of the \Pphi\ meson.
Positron and electron beams of equal energy collide at an angle 
of $\pi-25\mrad$, producing \Pphi\ mesons with a small momentum in the
horizontal plane: $p_\phi \approx 13\MeV /c$. \Pphi\ mesons decay 
$\sim 34\%$ of the time into nearly collinear \koko\ pairs. 
Because $J^{PC}(\phi)=1^{--}$, the kaon pair is in an antisymmetric state, 
so that the final state is always \ks\kl. The contamination from \kl\kl\ 
and \ks\ks\ final states is neglegible.
Therefore, the detection of a \kl\ signals the presence of a \ks\ of known
momentum and direction, independently of its decay mode. 
This technique is called \ks\ {\it tagging}. 
The analyzed sample corresponds to an integrated luminosity of $\sim 1.3\Lfb$,
 yielding $\sim 1.4$ billion of \ks\kl\ pairs. 
\begin{figure}[h!]
  \begin{center}    
    \includegraphics[totalheight=8.cm]{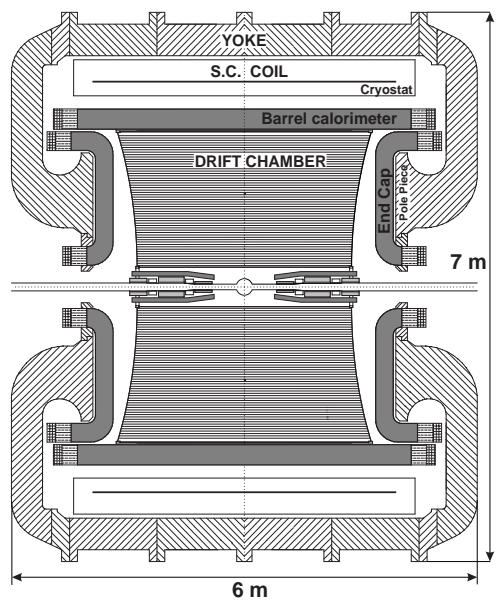}
    \caption{Vertical cross section of the KLOE detector.}
    \label{fig:kloedetector}
  \end{center}
\end{figure}
The \kloe detector (Fig.~\ref{fig:kloedetector}) consists of a 
large cylindrical drift chamber (DC), surrounded by a lead/scintillating-fiber
sampling calorimeter (EMC). A superconductig coil surrounding the calorimeter
provides a $0.52\T$ magnetic field. The drift chamber \cite{drift}, which is $4\m$
in diameter and $3.3\m$ long, has 12582 all-stereo tungsten sense wires and
37746 aluminium field wires. The chamber shell is made of carbon-fiber/epoxy
composite, and the gas used is a $90\%$ helium, $10\%$ isobutane mixture.
These features maximize trasparency to photons and reduce \kl$\to$\ks\
regeneration and multiple scattering. The DC position resolutions are 
$\sigma_{xy} \approx 150\um$ and $\sigma_{z} \approx 2\mm$.
DC momentum resolution is $\sigma(p_\perp)/p_\perp \approx 4\%$. Vertices
are reconstructed with a spatial resolution of $\sim3\mm$.

The calorimeter \cite{calo} is divided into a barrel and two endcaps,
contains a total of 88 modules, and covers $98\%$ of the solid angle.
The modules are read out at both ends by photomultiplier tubes. The arrival
times of particles and the three-dimensional positions of the energy deposits
are determined from the signals at the two ends. The read-out granularity
is $\sim 4.4\times 4.4 \cm ^2$; fired ``cells'' close in space and time
are arranged into a 
``calorimeter cluster''. For each cluster, the energy $E_{cl}$ is the
sum of the cell energies, and the time $t_{cl}$ and the position 
$\mathrm{r}_{cl}$ are calculated as energy-weighted averages over the fired 
cells. The energy and time resolutions are $\sigma_E/E = 5.7\%/ \sqrt{E(\GeV)}$
and $\sigma_t=57\ps / \sqrt{E(\GeV)}\oplus 100\ps$, respectively.

The calorimeter trigger \cite{ctrig} requires two local energy deposits 
above a threshold of $50\MeV$ in the barrel and
$150\MeV$ in the endcaps. 
Recognition and rejection of cosmic-ray events is also performed at 
the trigger level: events with two energy deposits above a $30\MeV$
threshold in the outermost calorimeter plane are rejected as cosmic-ray
events. Moreover, to reject residual cosmic rays and machine background
events, an offline software filter (FILFO) exploits calorimeter and DC
informations before tracks are reconstructed \cite{filfo}.
The trigger has a large time spread with respect to the beam crossing
time. However, it is synchronized with the machine RF divided by 4, 
$T_{sync} \sim 10.8\ns$, with an accuracy of $50\ps$.
An estimate of the time of the bunch crossing producing an event is 
determined offline during event recostruction. This value is subtracted
from the measured cluster times to obtain particle time-of-flight
(TOF) measurements.

The response of the detector to the decays of interest and the various
backgrounds were studied by using the \kloe\ Monte Carlo (MC) simulation
program~\cite{filfo}. Changes in the machine operation and background conditions
are simulated on a run-by-run basis.
The most important parameters are the beam
energies and the crossing angle, which are obtained from the analysis of 
Bhabha scattering events with $e^\pm$ polar angle above 45 degrees.
The average value of the center-of-mass energy is evaluated with a precision
of $30\keV$ for each $100\Lnb$ of integrated luminosity.
To study the background rejection, a MC sample of \Pphi\ decays to all possible 
final states has been used, equivalent to an integrated luminosity of $\sim 1.3\Lfb$.
A MC sample of $\sim 45000$ signal events has been also produced, to measure the analysis efficiency.

\section{Data analysis}
\label{sec:ana}

\subsection{\ks\ tagging}
\label{sec:kstag}

The identification of \kl-interaction in the EMC is used to tag the presence
of \ks\ mesons. The mean decay lenghts of \ks\ and \kl\ are 
$\lambda_S \sim 0.6\cm$ and $\lambda_L \sim 350\cm$, respectively. About $50\%$
of \kl's therefore reach the calorimeter before decaying. The \kl\ interaction 
in the calorimeter barrel (\kcr) is identified by requiring a cluster of energy greater than $125 \MeV$
not associated with any track, and whose time 
corresponds to a velocity $\beta = r_{cl}/ct_{cl}$ compatible with the kaon
velocity in the $\phi$ center of mass, $\beta^* \sim 0.216$, after the 
residual $\phi$ motion is considered. Cutting at $0.17 \le \beta^* \le 0.28$ we selected 
$\sim 450$ million \ks-tagged events (\kcr\ events in the following), which are used as a starting sample for the 
\DKSee\ search.

\subsection{Signal preselection and background normalization}
\label{sec:presele}

\begin{figure}[h!]
  \begin{center}    
    \includegraphics[totalheight=8.5cm]{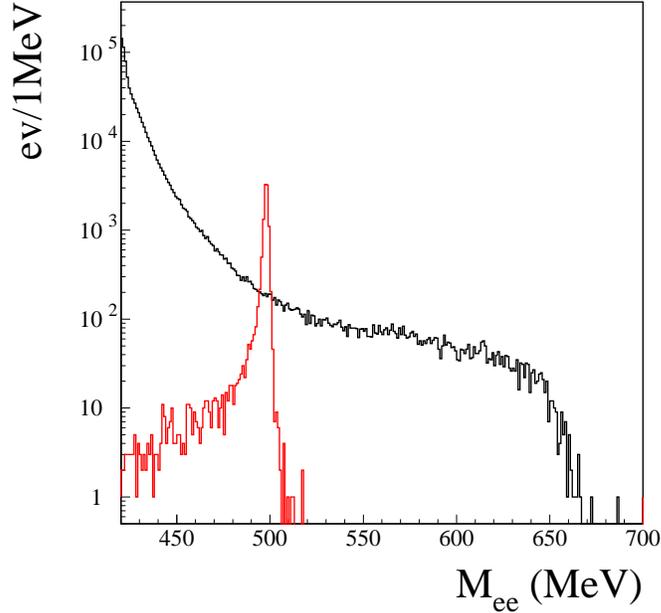}
    \caption{Two-track invariant mass evaluated in electron hypothesis for MC signal (red) and 
background (black) events.}
    \label{fig:minvee}
  \end{center}
\end{figure}

\begin{figure}[h!]
  \begin{center}    
    \includegraphics[totalheight=8.5cm]{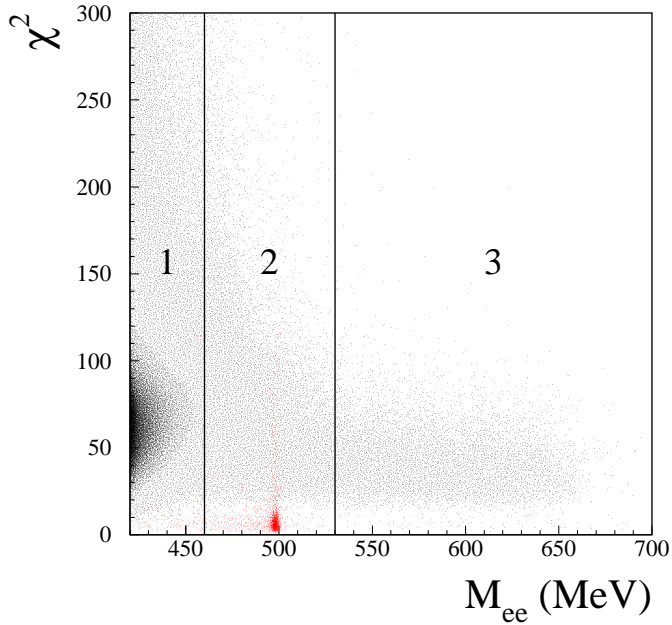}
    \caption{Scatter plot of $\chi^2$ versus $M_{ee}$ for MC signal (red) and background events (black)}
    \label{fig:chi2minv}
  \end{center}
\end{figure}

\begin{figure}[h!]
  \begin{center}    
    \includegraphics[totalheight=6.5cm]{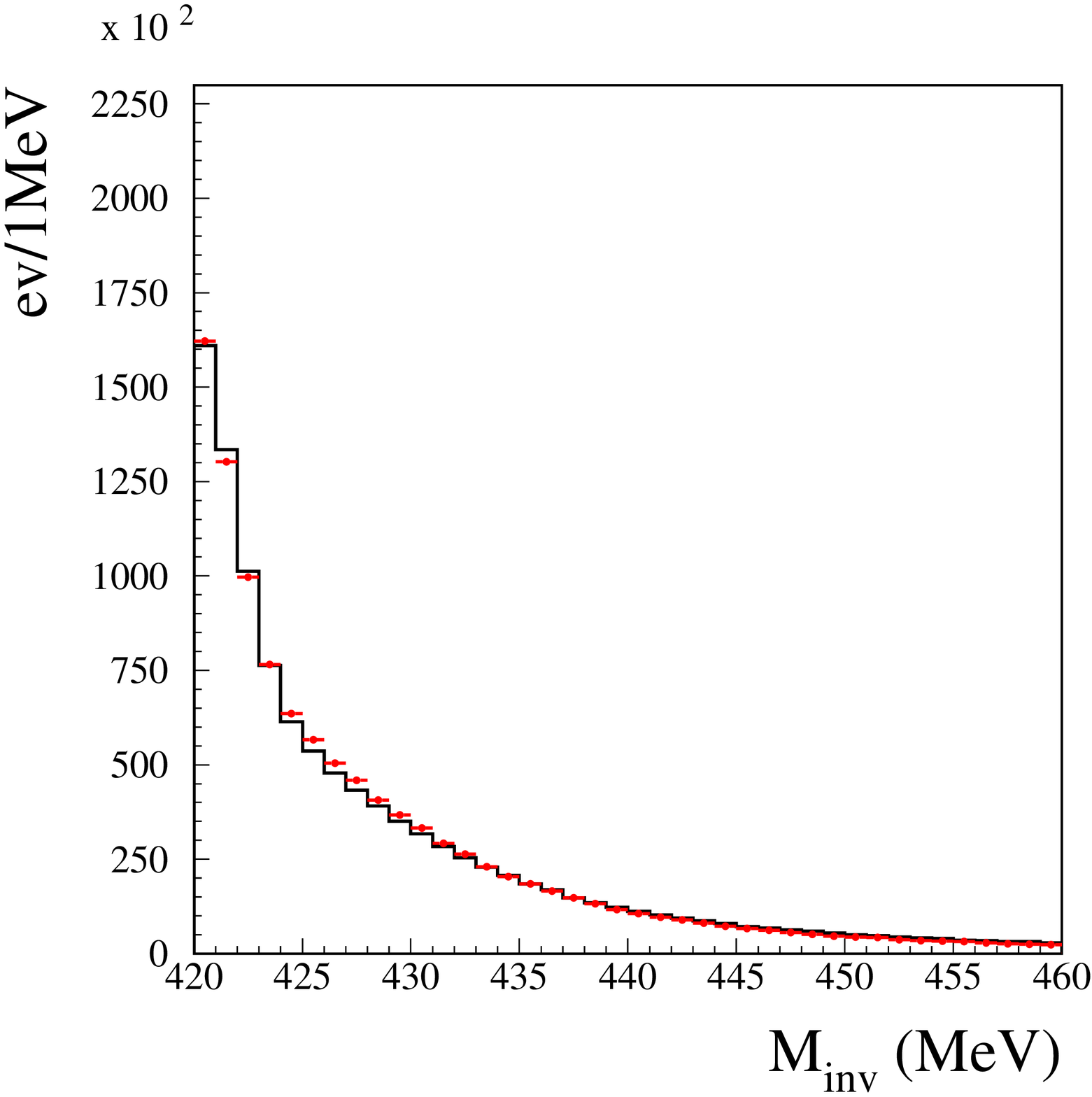}
    \includegraphics[totalheight=6.5cm]{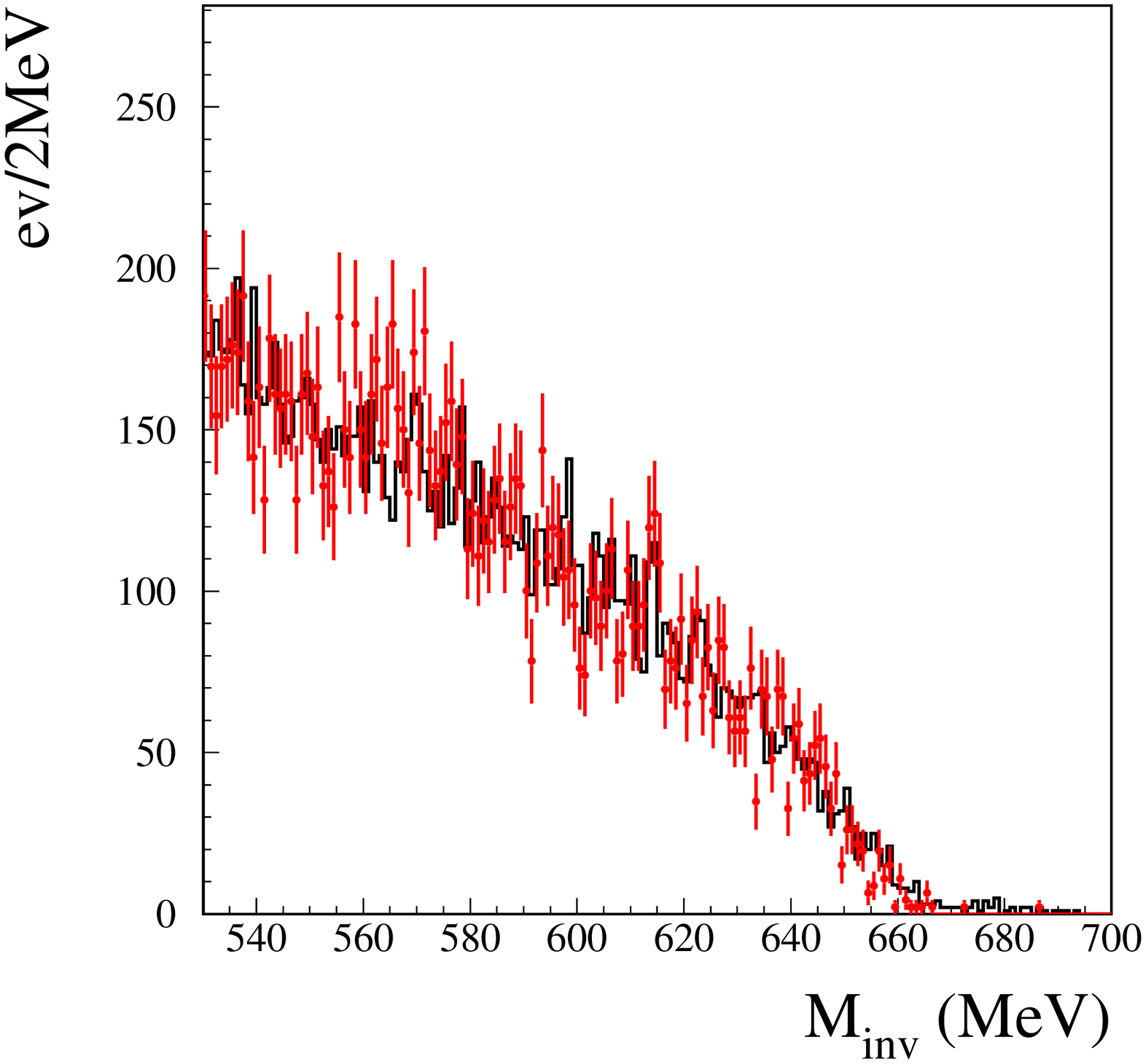}
    \caption{Data-MC comparison for $M_{ee}$ spectra in region 1 (left) and region 3 (right), after normalization:
data are represented by the black solid line, MC by red points.}
    \label{fig:sidebands}
  \end{center}
\end{figure}

\DKSee\ events are selected by requiring the presence of two tracks of opposite 
charge with their point of closest approach to the origin inside a cylinder $4\cm$ in radius and 
$10\cm$ in length along the beam line. Moreover, the two tracks are required to form a vertex with position in the transverse plane 
$\rho <  4 \cm$.
The track momenta and polar angles must satisfy the fiducial cuts 
$120\MeV /c \le p \le 350\MeV /c$ and 
$30^{\circ} \le \theta \le 150^{\circ}$. The tracks must also reach the EMC
without spiralling, and have an associated cluster.  
In Fig.~\ref{fig:minvee}, the two-track invariant mass evaluated in electron hypothesis ($M_{ee}$) is shown for both 
MC signal and background samples. A preselection cut requiring $M_{ee}> 420 \MeV/c^2$ has been applied, which rejects most
of $\DKSpippim$ events, for which $M_{ee}\sim 409 \MeV/c^2$. The residual background has two main components: \DKSpippim\ events,
populating the low $M_{ee}$ region, and \Dphipippimpio\ events, spreading over the whole spectrum. The \DKSpippim\ events
have such a wrong reconstructed $M_{ee}$ because of track momentum resolution or one pion decaying into a muon. The \Dphipippimpio\ 
events enter the preselection because of a machine background cluster, accidentally satisfying the \kcr\ algorithm. 
After preselection we are left with $\sim 10^6$ events.
To have a better separation between signal and background, a $\chi^2$-like 
variable is defined, collecting informations from the clusters associated
to the candidate electron tracks.
Using the MC signal events we built likelihood functions based on:
\begin{itemize}
\item the sum and the difference of $\delta t$ for the two tracks, where $\delta t = t_{cl} - L/\beta c$ is evaluated in 
electron hypothesis;
\item the ratio $E/p$ between the cluster energy and the track momentum, for both charges;  
\item the  cluster centroid, for both charges. 
\end{itemize}
In Fig.~\ref{fig:chi2minv}, the scatter plot of $\chi^2$ versus $M_{ee}$ is shown, for MC signal and background
events. The $\chi^2$ spectrum for background is concentrated at higher values respect to signal, since both 
\DKSpippim\ and \Dphipippimpio\ events have pions in the final state.
To assess the MC background normalization, two sidebands are defined in the invariant mass: $M_{ee}<460\MeV/c^2$ 
(region 1 in Fig.~\ref{fig:chi2minv}), and $M_{ee}>530\MeV/c^2$ (region 3 in Fig.~\ref{fig:chi2minv}).
\DKSpippim\ events largerly dominate on \Dphipippimpio\ in region 1, the opposite occurring in region 3. In the two 
sidebands, a normalization factor is evaluated for each background component, by fitting the MC spectra to data. A comparison between
data and MC after the fit is shown in Fig.~\ref{fig:sidebands}. 

\subsection{Background rejection}
\label{sec:bkgrej}
\begin{figure}[h!]
  \begin{center}    
    \includegraphics[totalheight=6.cm]{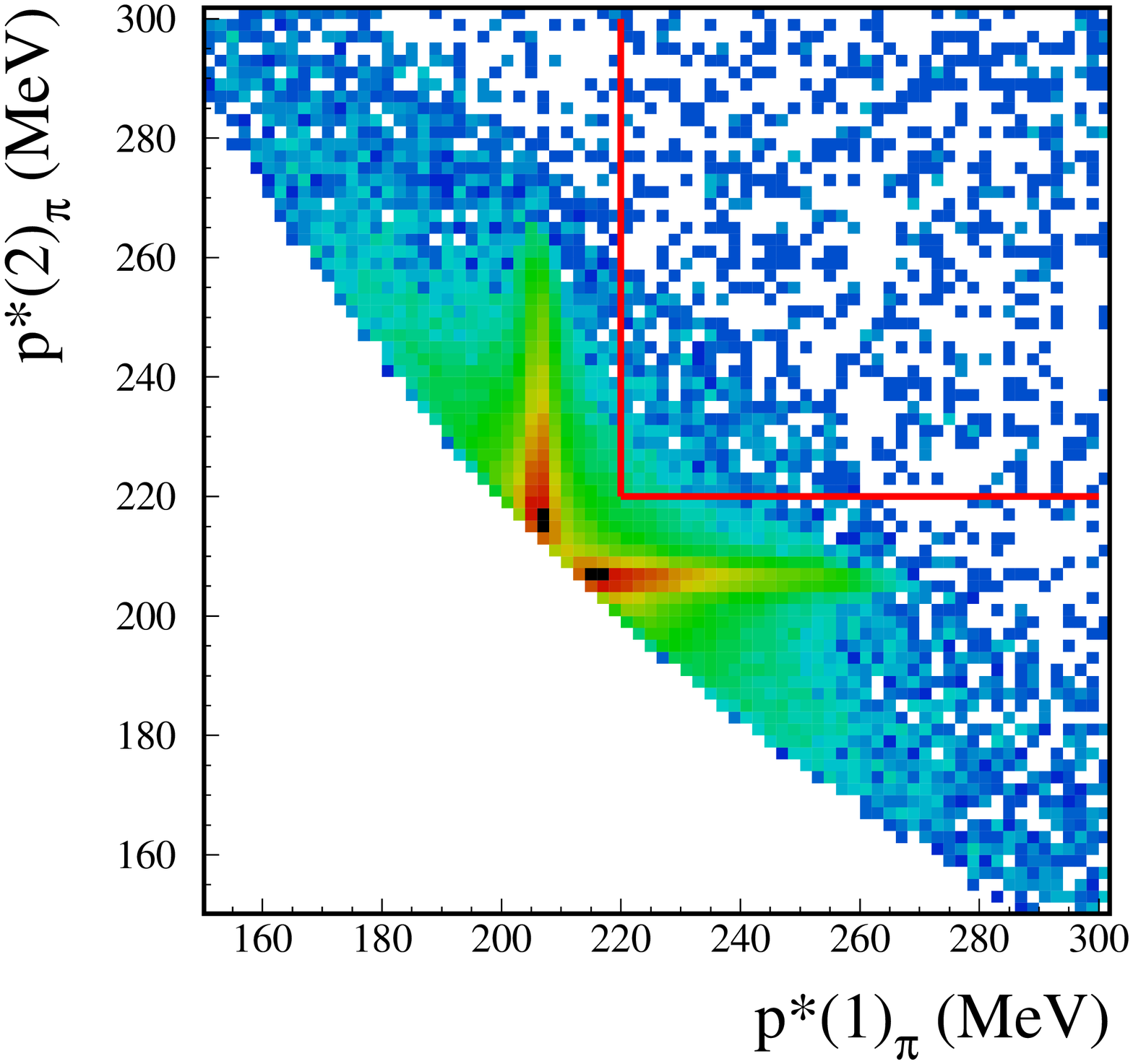}
    \includegraphics[totalheight=6.cm]{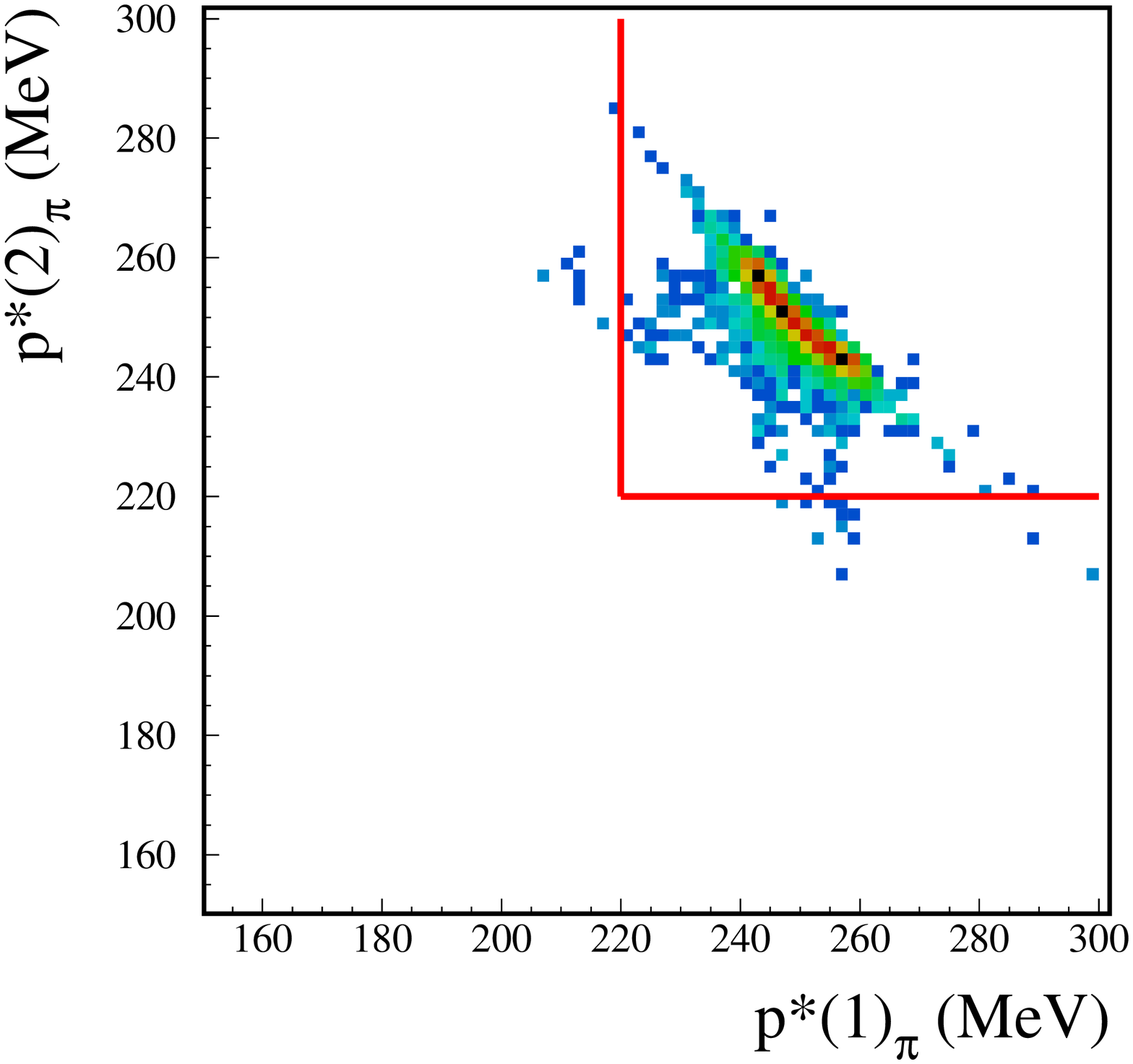}
    \caption{Scatter plot of track momenta in the \ks\  rest frame ($p^*_\pi$), for MC background in region 1 (left), 
     and for MC signal (right).}
    \label{fig:pstar}
  \end{center}
\end{figure}

\begin{figure}[h!]
  \begin{center}    
    \includegraphics[totalheight=6.cm]{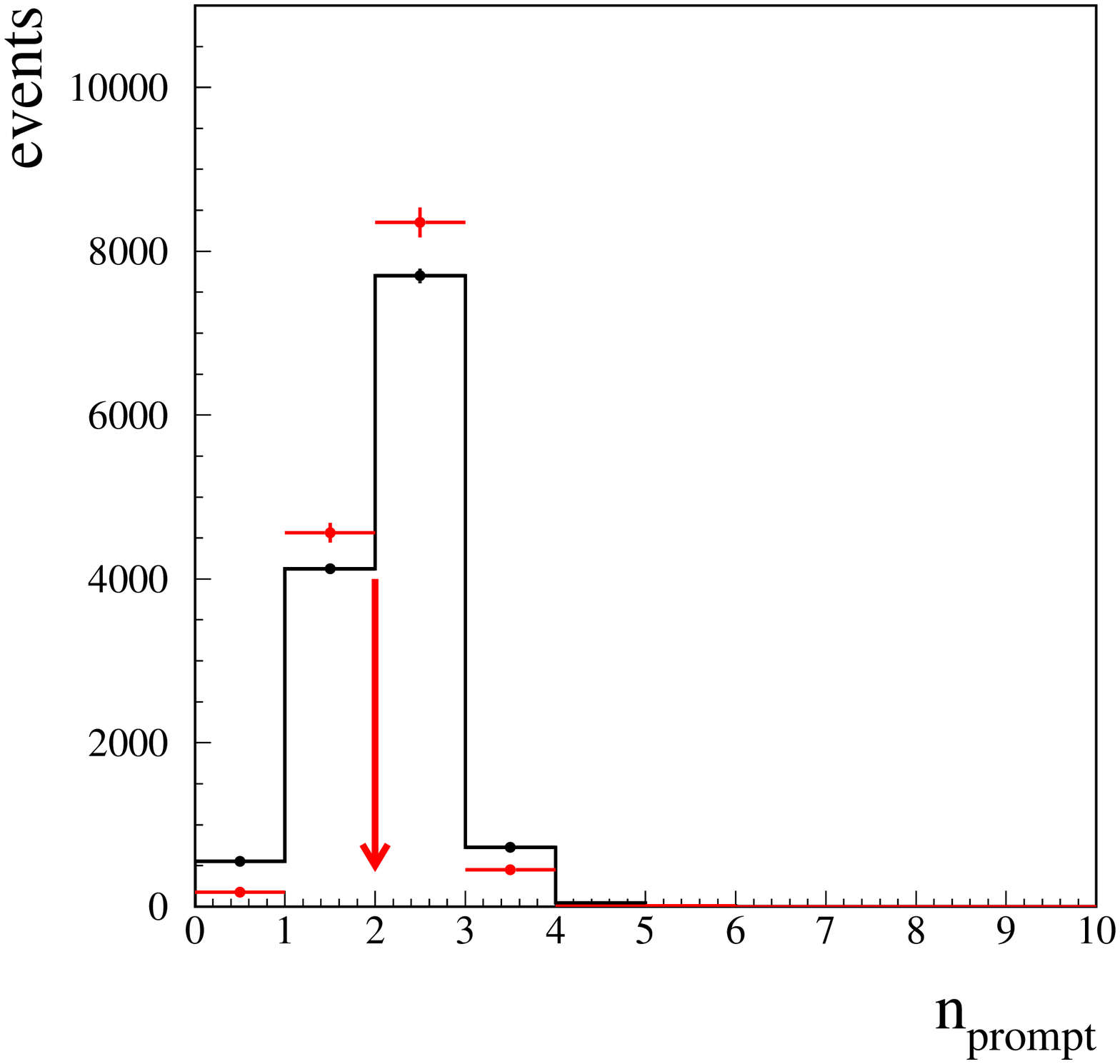}
    \includegraphics[totalheight=6.cm]{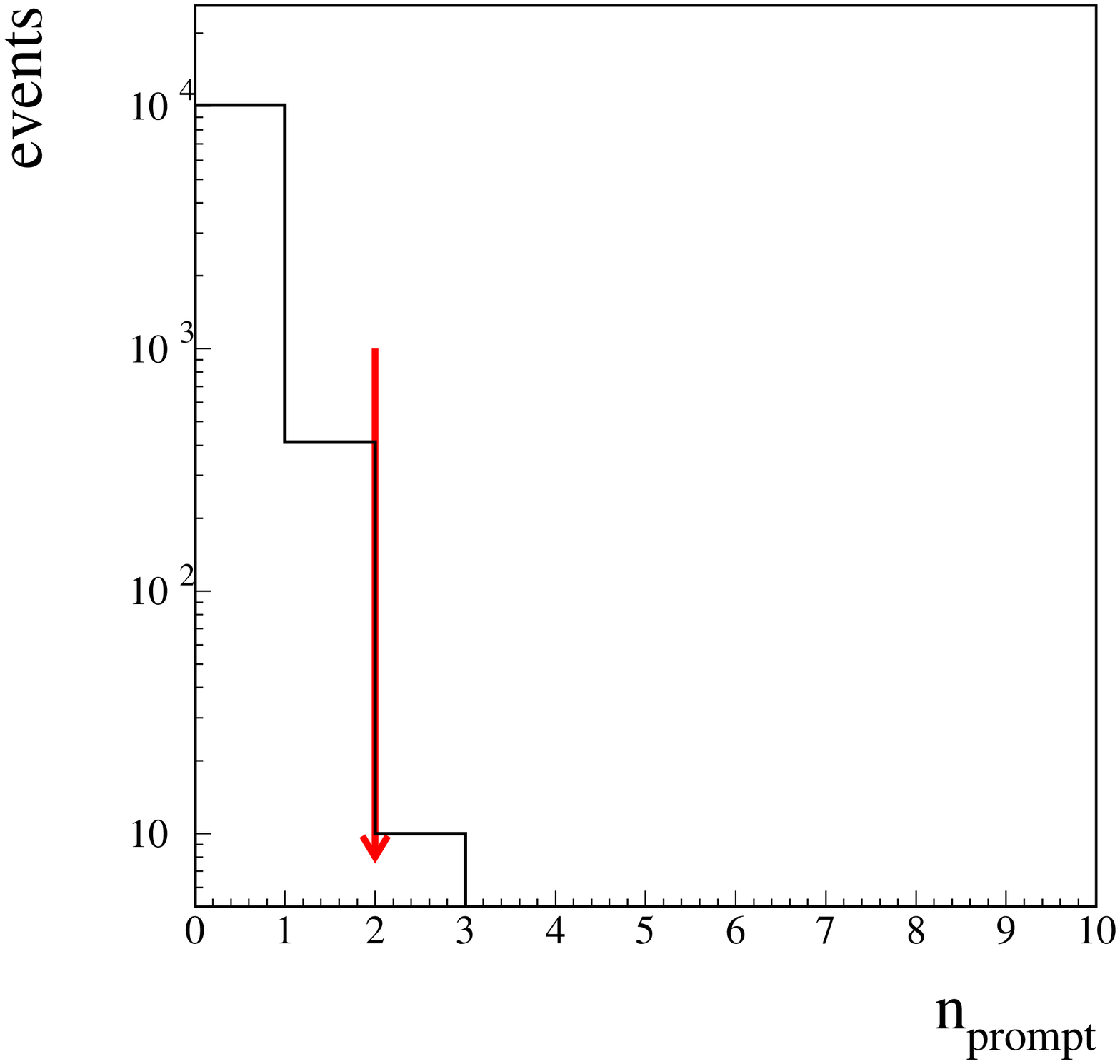}
    \caption{Number of prompt photons ($N_{prompt}$) for data (black) and MC background (red) in region 3 (left), and for MC signal (right).}
    \label{fig:nprompt}
  \end{center}
\end{figure}

\begin{figure}[h!]
  \begin{center}    
    \includegraphics[totalheight=6.cm]{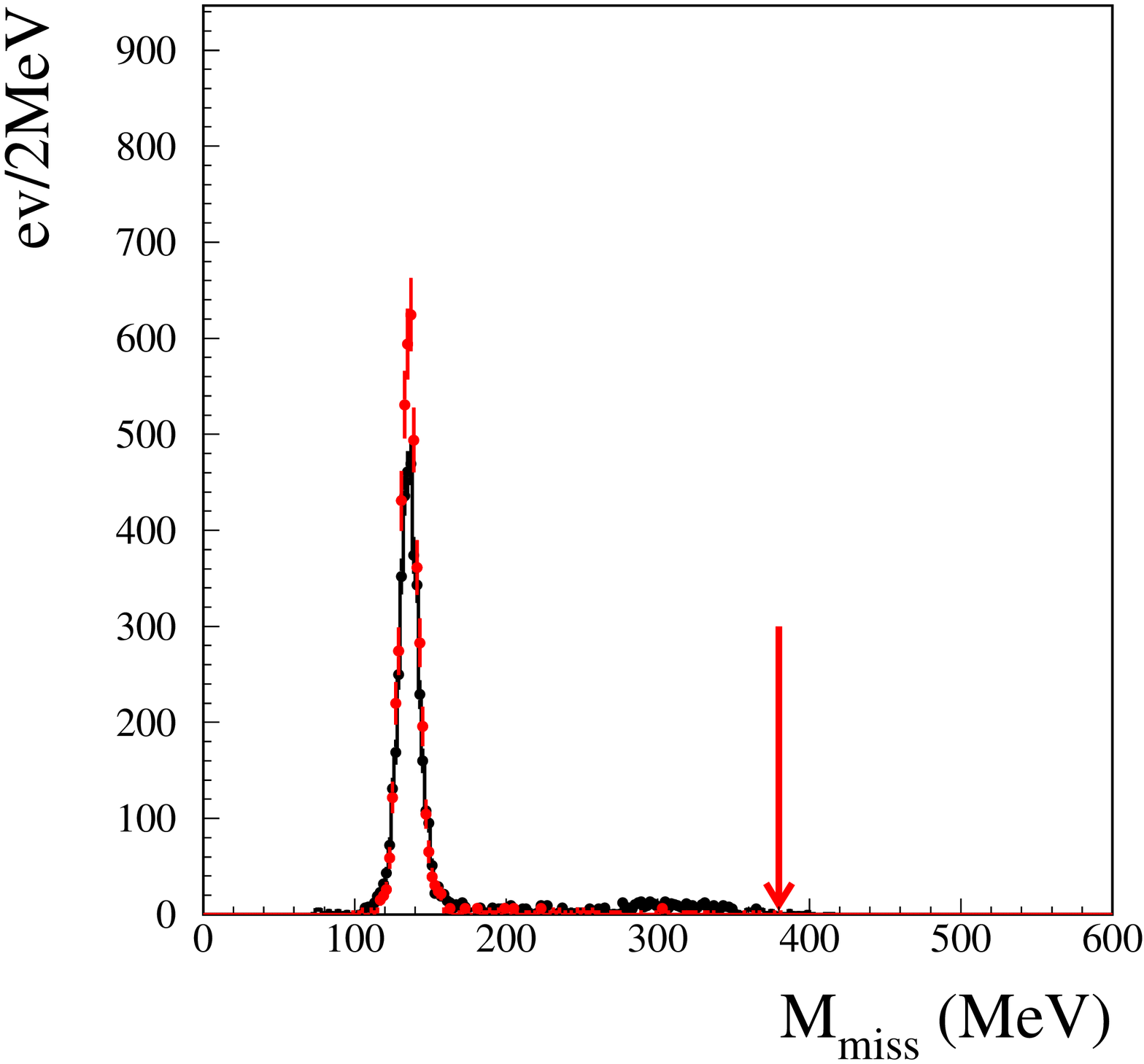}
    \includegraphics[totalheight=6.cm]{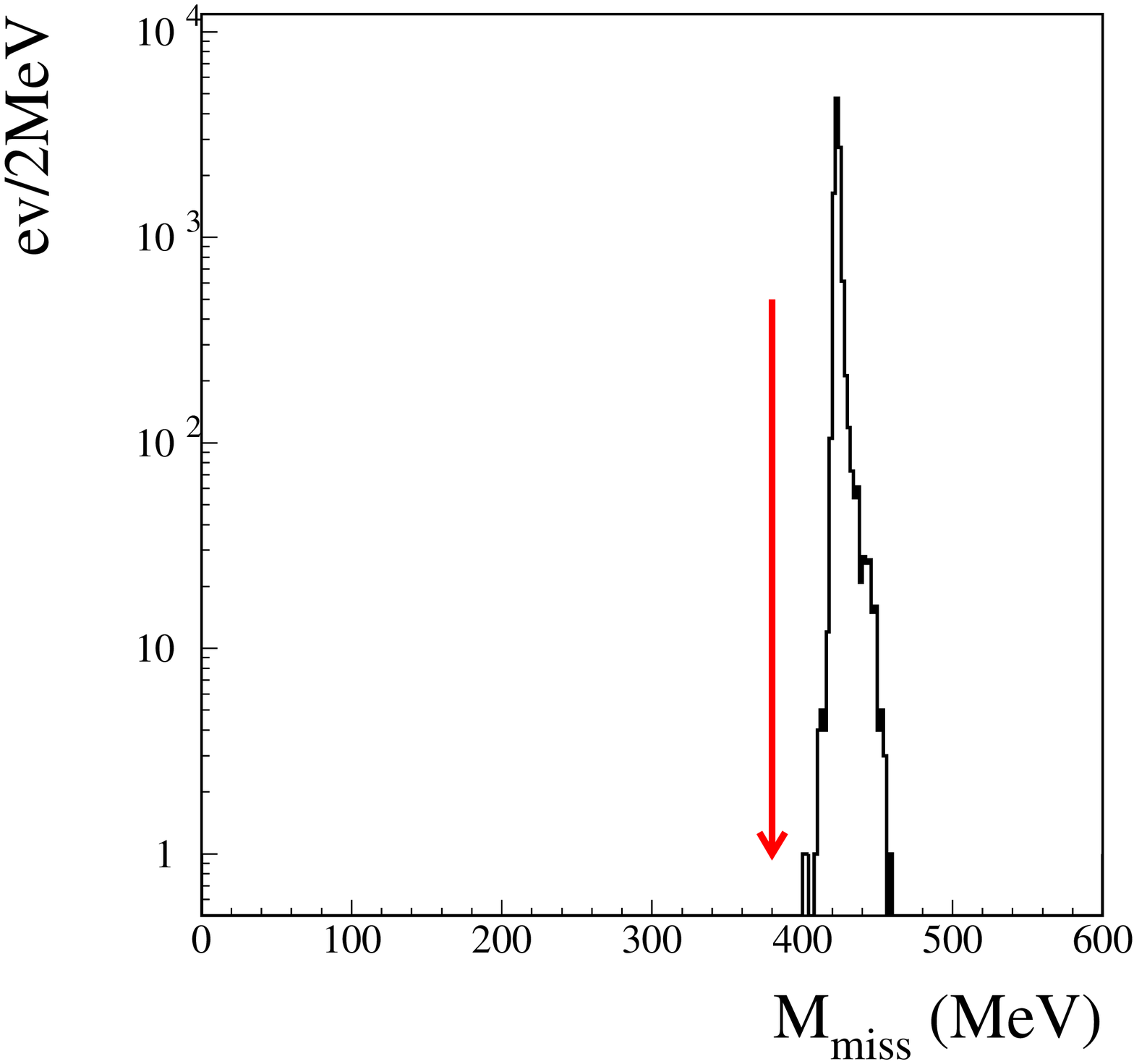}
    \caption{Missing mass ($M_{miss}$) for data (black) and MC background (red) in region 3 (left), and 
      for MC signal (right).}
    \label{fig:mmiss}
  \end{center}
\end{figure}

A signal box to select the \DKSee\ events can be conveniently defined in the region 2 of the $M_{ee}-\chi^2$ plane 
(see Fig.~\ref{fig:chi2minv}); nevertheless we investigated some more independent requirements in order to reduce the background 
contamination as much as possible before applying the $M_{ee}-\chi^2$ selection. These cuts have been tuned on the $M_{ee}$ sidebands, 
which are also used to check data-MC consistency after each step of the analysis, and they are summarized below.
Charged pions from \DKSpippim\ decay have a momentum in the \ks\ rest frame $p_\pi^{\ast}\sim 206\MeV/c$.
The distribution of track momenta in the \ks\ rest frame, evaluated in the pion mass hypothesis, is shown in Fig.~\ref{fig:pstar}, for MC 
background in region 1, and for MC signal.  
For most of \DKSpippim  decays, at least one pion has well reconstructed momentum, so that  
the requirement $min(p^{\ast}_\pi(1), p^{\ast}_\pi(2))  \ge 220 \MeV/c$ rejects $\sim 99.8\%$ of these events, while retaining $\sim 97\%$ of the
signal.  
To reduce the \Dphipippimpio\ background, we can use the fact these events have two photons coming from the interaction point.  
The distribution of the number of prompt photons ($\delta t = t_{cl} - r_{cl}/c < 5\sigma_t$) for data and MC background in region 3,
 and for MC signal, is shown in Fig.~\ref{fig:nprompt}. We require $N_{prompt}\le 1$, thus rejecting $\sim 65\%$ of background and 
only $\sim 0.1\%$ of signal events.
Further rejection on \Dphipippimpio\ events is achieved by cutting on the missing mass, evaluated as
$M_{miss}^2 = \left(\tilde{P}_\phi-\tilde{p}_+ -\tilde{p}_- \right)^2$,
where $\tilde{P}_\phi$ is the $\Pphi$ four-momentum and $\tilde{p}_\pm$ are the charged track four-momenta, in pion hypothesis. 
Distribution of $M_{miss}$ is shown in Fig.~\ref{fig:mmiss}, for data and MC background in region 3, and for MC signal. 
We cut at $M_{miss}>380 \MeV/c^2$, rejecting almost completely the 3-pion background, peaking at the $\pi^{0}$ mass. 
A comparison between data and MC counts on the $M_{ee}$ sidebands after each cut is shown 
in Tab.~\ref{tab:sidebands}, which demonstrates the reliability of the 
background simulation. 
\begin{table}[h!]
\begin{center}
\begin{tabular}{|c|ccc|ccc| }
\hline
 & \multicolumn{3}{c|}{Region 1} & \multicolumn{3}{c|}{Region 3}\\
Cut & Data & MC bkg & $\Delta/\sigma$ & Data & MC bkg & $\Delta/\sigma$\\ \hline  
$p^*_\pi$   & 6061  &  6285(125)   & -1.5  & 13144  & 13590(184)  & -2.0  \\  
$N_{prompt}$&  2778 &  2982(80)   & -2.1  & 4672   & 4745(103)   & -0.6  \\  
$M_{miss}$  &  1370 & 1407(53)    &  -0.6 & 20     & 5(3)       & 2.8  \\ \hline 
\end{tabular}
\caption{Data and MC counts in the $M_{ee}$ sidebands after each step of the background rejection; the relative difference between 
data and MC is also reported.}
\label{tab:sidebands}
\end{center}
\end{table}

\subsection{Signal box definition}
\label{sec:sbox}
A signal box is defined in $M_{ee}-\chi^2$ plane. To achieve the best background rejection, with maximum signal 
efficiency, we scan over a large set of $M_{ee}-\chi^2$ cut configurations.  
This optimization procedure is based on MC only, without looking at the number of observed events ($N_{obs}$) on data.
The best configuration is:
\begin{equation}
\left\{ \begin{array}{l}
492\MeV \le M_{ee} \le 504\MeV\\
\chi^2\le 20
\end{array} \right.
\end{equation}
Applying this selection to the data sample we obtain $N_{obs}=3$.
The expected background estimated from MC is $\mu_{B}=7.1\pm 3.6$, which 
takes into account the MC statistics and uncertainty on the normalization factors. Using a bayesian
approach~\cite{Helene}, we evaluate the upper limit on the expected number of signal
events $\mu_S$ to be $UL(\mu_S) = 4.3$, at $90\%$ CL.

\subsection{Radiative corrections}

Given the chosen invariant mass selection, we actually measure the upper limit 
on $\DKSee(\gamma)$, with $E_\gamma^*<6\MeV$. Two processes are  
expected to contribute to photon emission, not interferring with each other:
\begin{itemize}
\item the inner bremsstrahlung photon emission, $\DKSee+\gamma_{IB}$;
\item a \DKSgg\ decay, with one photon conversion, $\ks\to\gamma\gamma^*\to\gamma e^+ e^-$.
\end{itemize}  
For the first process we estimate a correction given by~\cite{Gatti}:
$$
\epsilon_{rad} = \frac{\Gamma(\DKSee(\gamma)\, , \, E_\gamma^*<6\MeV)}{
\Gamma(\DKSee(\gamma))} = 0.8
$$
The second process is strongly peaked around $M_{ee}\sim 2m_e$~\cite{Bergstrom}, and a 
$BR(\ks\to\gamma\gamma^*\to\gamma e^+ e^-)\sim 2\times 10^{-16}$ is expected in our final mass window.
So, our result consists in a limit on the IB emission, being insensitive, in this $M_{ee}$ range,
to the photon conversion process.

\section{Results}
\label{sec:re}

The total selection efficiency on \DKSee\ events is evaluated by MC, using the following parametrization:
$$
\epsilon_{sig} = \epsilon(K_{crash}) \times \epsilon(sele \vert K_{crash}), 
$$
where $ \epsilon(K_{crash})$ is the tagging efficiency, and $ \epsilon(sele \vert K_{crash})$ is the signal selection
 efficiency on the sample of tagged events.  
The number of \DKSpippim\ events $N_{\pi^+\pi^-}$ counted on the same sample of \ks\ tagged events
is used as normalization, with a similar expression for the efficiency. 
The upper limit on BR(\DKSee) is evaluated as follows:
\begin{eqnarray}
 UL(BR(\DKSee))& = & \nonumber \\
 UL(\mu_s)     &\times & \mathcal{R}_{tag}\times 
\frac{\epsilon_{\pi^+\pi^-}(sele \vert K_{crash})}{
\epsilon_{sig}(sele \vert K_{crash})}\times
\frac{BR(\DKSpippim)}{N_{\pi^+\pi^-}}, \nonumber 
\end{eqnarray}
where $\mathcal{R}_{tag}$ is the tagging efficiency ratio, corresponding to
a small correction due to the \kcr\ algorithm dependence on \ks\ decay mode.
Determination of both $\mathcal{R}_{tag}$ and  $\epsilon_{\pi^+\pi^-}(sele \vert K_{crash})$ 
is discussed in detail in Ref.~\cite{kspipi}.
Using as input values:
\begin{itemize} 
\item  $UL(\mu_S) = 4.3$, at $90\%$ CL.
\item $\epsilon_{sig}(sele \vert K_{crash}) = 
  \underbrace{0.697}_{cuts}\  \times\underbrace{0.8}_{rad} = 0.558(4)$
\item $\mathcal{R}_{tag} = 0.9634(1)$
\item $\epsilon_{\pi^+\pi^-}(sele \vert K_{crash}) = 0.6102(5)$
\item $N_{\pi^+\pi^-} = 148,184,688$
\end{itemize}
we obtain
$$
UL(BR(\DKSee(\gamma))) =  2.1 \times 10^{-8}, \; {\rm at} \;90\%\,{\rm CL}.
$$
Systematics uncertanties, related to background normalization, are at the level of 2\%.
Our measurement improves by a factor of $\sim 7$ on the
 CPLEAR result~\cite{cplearksee}, for the first time including radiative
corrections in the evaluation of the upper limit.

\bibliographystyle{elsart-num}

\end{document}